\documentstyle[12pt,titlepage]{article}

\font\grande=cmr10 scaled \magstep4
\font\medio=cmr10 scaled \magstep2
\outer\def\beginsection#1\par{\medbreak\bigskip
      \message{#1}\leftline{\bf#1}\nobreak\medskip
\vskip-\parskip
      \noindent}

\def\laq{\raise 0.4ex\hbox{$<$}\kern -0.8em\lower 0.62
ex\hbox{$\sim$}}
\def\gaq{\raise 0.4ex\hbox{$>$}\kern -0.7em\lower 0.62
ex\hbox{$\sim$}}
\def\beq{\begin{equation}}
\def\eeq{\end{equation}}
\def\bea{\begin{eqnarray}}
\def\eea{\end{eqnarray}}
\def\bean{\begin{eqnarray*}}
\def\eean{\end{eqnarray*}}

\begin{document}
\bibliographystyle {unsrt}

\titlepage
\begin{flushright}
\vspace{15mm}
CERN-TH/2001-278 \\
hep-th/0110129 \\

\end{flushright}
\vspace{15mm}
\begin{center}
{\grande Large-N bounds on, and  compositeness limit of,
 gauge and gravitational interactions}\\

\vspace{15mm}

  G. Veneziano 
\vspace{6mm}

{\sl Theory Division, CERN, CH-1211 Geneva 23, Switzerland} \\
and  \\
{\sl Laboratoire de Physique Th\`eorique, Universit\`e Paris Sud, 91405, Orsay, France} \\
\end{center}

\vskip 2cm
\centerline{\medio  Abstract}
\vskip 5mm
\noindent
In a toy model of gauge  and gravitational interactions
in $D \ge 4$ dimensions, endowed with an invariant
UV cut-off $\Lambda$, and containing 
a large number $N$ of non-self-interacting
matter species, 
the physical gauge and gravitational couplings at the cut-off,
$\alpha_g \equiv g^2 \Lambda^{D-4}$ and
$\alpha_G \equiv G_N \Lambda^{D-2}$,
are shown to be bounded by appropriate powers  of ${1\over N}$. 
This  implies that
the infinite-bare-coupling 
 (so-called compositeness) limit of these theories
 is smooth, and can even resemble
our world.
We argue that
such a result, when extended to more realistic situations, 
 can help avoid large-$N$ violations of entropy bounds, 
solve the dilaton stabilization and GUT-scale problems in superstring 
theory, and provide a new possible candidate for quintessence.
\vspace{5mm}

\vfill
\begin{flushleft}
CERN-TH/2001-278\\
October 2001\\

\end{flushleft}

\newpage
\section{ The toy model and the claim}
Consider a  toy model of gauge and gravitational
interactions in $D \ge 4$ space-time dimensions,
 minimally coupled to a large number 
of  spin $0$ and spin  $1/2$ matter fields. 
Let us endow the model with a cut off $\Lambda$,
assumed to be finite and to
preserve gauge invariance and general covariance. The toy model
is supposed to mimic a bona-fide higher dimensional
UV-finite theory of all interactions
such as those provided by superstring theory.
Let us also neglect, for the moment,  matter self interactions.
The tree-level action of the model thus reads, in obvious notations
\bea
S_{0} &=& - {1 \over 2} \int d^{D}x \sqrt{-g} 
\left[ \kappa_{0}^{-2} R + {1 \over 2} g_{0}^{-2} 
\sum_{k=1}^{N_1}  
F^{k}_{\mu\nu} F^{k\mu\nu} \right] 
+  \sum_{i=1}^{N_{0}}  
\left(  (D_{\mu} \phi)_{i} (D^{\mu} \phi)_{i}
 + m_0^2 \phi_{i}^2 \right) \nonumber \\
      &+& \int d^{D}x \sqrt{-g} \left[ \sum_{j=1}^{N_{1/2}} \left( \bar{\psi}_{j} 
(i\gamma \cdot D \psi)_{j}
 + m_{1/2} \bar{\psi}_{j} \psi_{j} \right) 
 + ~ \dots 
\right] ,
\label{treeaction}
\eea
where  dots stand for an ultraviolet completion of the model
 implementing the UV cutoff. We have given for simplicity a 
common mass $m_0$ to all
the spin zero fields and a common mass $m_{1/2}$ to all spin $1/2$ fields.
These masses are assumed to be small compared to the UV cut-off $\Lambda$.

We are interested in the value of the renormalized 
gauge and gravitational couplings, $g^2$ and 
$\kappa^{2} = 8 \pi G_{N}$, as a function of their
bare values, $g_0^2$ and $\kappa_0^{2}$, and of $\Lambda$, when the total number 
of matter fields $N = N_{0} + N_{1/2} \rightarrow \infty$, 
while  their  relative ratios are kept fixed.
We claim that, in the above-defined model, and modulo
a certain generic assumption  about one-loop contributions, the 
following bounds hold:
\bea
\alpha_g \equiv g^2 \Lambda^{D-4} &<& {c_1 \over N^p} \; , D > 4\;,
 \nonumber \\
\alpha_G \equiv G_N \Lambda^{D-2} &<&  {c_{2} \over N} \; ,  D \ge 4\;,
\label{bound}
\eea
where $c_{1}, c_{2}$ are positive constants (typically  smooth functions
of the relative abundances $N_{i}/N$) to be computed 
at the one-loop level, and $p$ is a number between
$0$ and $1$. The case of the gauge coupling at $D=4$
 needs a separate discussion because of infrared effects.

We also claim that both bounds are saturated in the compositeness 
(infinite-bare-coupling) limit and, therefore, that such a limit exists, is smooth,
and can possibly be a realistic one, in agreement with  older proposals \cite{CMPP}.

\section{Proving the claim}

Consider the Feynman  path integral corresponding to the action 
(\ref{treeaction}) and integrate out completely (i.e. on all
scales) the matter fields. Since they 
only appear quadratically the exact result is:
\bea
I &=& \int  dg_{\mu\nu} d A_{\mu}^k d \phi_{i} d \psi_j
 {\rm exp} \left(i (S_{0, {\rm gravity}} +  S_{0, {\rm gauge}} +
 S_{0, \rm matter} + \dots) \right) \nonumber \\
  &=& 
\int  dg_{\mu\nu}  d A_{\mu}^k {\rm exp} (i S_{{\rm eff}}) \; , 
\label{Fintegral}
\eea
where
\beq
 S_{{\rm eff}} = S_{0, {\rm gravity}} + S_{0, {\rm gauge}}
 - 1/2  ~~ {\rm tr~log \nabla^2} (g, A) +  
  {\rm tr~log} (\gamma \cdot D (g, A))  
 \; , 
\label{Seff}
\eeq
and the trace includes the sum over the representations to which
the matter fields belong, in particular a sum over ``flavour" indices.

The latter two terms in $S_{{\rm eff}}$ can be evaluated by standard 
heat-kernel
techniques \cite{Vilko}. The result is well known to contain local as well 
as non-local terms. In  $D>4$ the non-local terms start with at least 
four derivatives. In $D=4$ this is still true for the gravity
part but non-local contributions appear already in the
$F^2$ terms of the gauge-field action. Thus we write:
\beq
  S_{{\rm eff}} =  S_{0, {\rm gravity}}\left(1 + 
(c_0 N_0 + c_{1/2} N_{1/2})  \kappa_0^2 \Lambda^{D-2} \right)
+  S_{0, {\rm gauge}} \left(1 + g_0^2 (\beta_0 + \beta_{1/2}) {\Lambda^{D-4} \over
D-4} 
\right) +  S' ,
\label{Seffexp}
\eeq
with the following explanatory remarks.
 $S'$ contains higher derivative (and generally non local) terms.
 The other terms in the  gauge plus gravity effective action are
local with the already mentioned exception of the gauge kinetic term
in $D=4$: this is indicated symbolically by the presence of a 
pole at $D=4$, to be explained better below.
Note that the corrections to the tree-level action
are independent of the bare gauge and gravitational couplings (the explicit
factors appearing in (\ref{Seffexp}) being canceled by those implicit in the
definition of the tree-level actions). Finally, the constants 
$c_0, c_{1/2}, \beta_0, \beta_{1/2}$ are in principle computable
 in any given theory; their order
of magnitude in the large $N$ limit will be discussed below.

We now have to discuss the effect of including gauge and gravity loops or,
if we prefer, to complete the functional integral by integrating over
$g_{\mu\nu}$ and $A_{\mu}^k$ after having introduced suitable
sources. This is, in general, quite non-trivial,
 however appropriate large-$N$ limits can help. Let us start with
the effect of these last integrations on the gauge kinetic term, i.e. with
the renormalization of the gauge coupling due to gravity and gauge loops.
Obviously, such a renormalization adds to the one due to matter loops, and
already included in (\ref{Seffexp}). If we consider a large-$N$ limit such
that, not only $N_0, N_{1/2} \rightarrow \infty$, but also
$\beta_0  + \beta_{1/2} \rightarrow \infty$, the effective coupling after
matter-loop renormalization is arbitrarily small. In this case, the 
one-gauge-loop contribution dominates the remaining functional integrals (the theory
having become almost classical)
 and we get for the final low-energy gauge effective action
\beq
\Gamma_{eff}^{gauge} = - {1 \over 4} \int \sqrt{-g} 
\left[  g_{0}^{-2} + (\beta_0  + \beta_{1/2})
 {\left(\Lambda^{2 \epsilon} -
(q^2 + m^2)^{\epsilon} \right)\over \epsilon} - 
\beta_1 {\left(\Lambda^{2 \epsilon} -
(q^2)^{\epsilon} \right)\over \epsilon}  \right]  F_{\mu\nu}^2 \;,
\label{Gammagauge}
\eeq
 where $\epsilon = (D-4)/2$ and, for the sake of notational simplicity,
 we have taken $m_0 = m_{1/2} = m$.

In order for our approximations to be justified we need to argue that
the quantity $(\beta_0  + \beta_{1/2})$ is sufficiently large and positive, 
indeed that it is parametrically 
larger than the gauge field contribution $\beta_1$. The latter
is proportional to the quadratic Casimir of the adjoint representation $C_A$.
Recalling the way matter fields couple to gauge fields, we
 find that this condition is satisfied
provided $\beta_0  + \beta_{1/2} \sim C_M N_f \frac{d_M}{d_A} \gg C_A$, 
where $d_M, C_M$ represent  dimensionality
and  quadratic Casimir for the matter representation $M$, respectively.  In a QCD-like theory
with gauge group $SU(N_c)$ this would correspond to a large $N_f/N_c$ ratio.

What happens to the effective theory at low energy depends very much on
whether $D$ is larger or equal to $4$.
If $D>4$, $\Gamma_{eff}^{gauge}$ is local and we can neglect the corrections
proportional to $m^2$ or $q^2$. The renormalized gauge coupling is bound, at all
scales, i.e.
\beq
4 \pi \alpha_g \equiv g^{2} \Lambda^{D-4} \sim [g_{0}^{-2} \Lambda^{4-D} + 
(\beta_0  + \beta_{1/2}) - 
\beta_1]^{-1} \le (\beta_0  + \beta_{1/2} - \beta_1 )^{-1} ,~ D>4 ,
\label{alphabound}
\eeq
and the upper bound is reached at infinite bare coupling, 
$g_{0}^{-2} \rightarrow 0$. Note that this conclusion only holds ``generically" i.e.
under the assumption that  no cancellation between $\beta_0$, $\beta_{1/2}$ and
$\beta_1$ prevents their combination appearing in (\ref{alphabound})
 from growing  large and positive (the relative signs appearing there are 
a matter of conventions, the actual signs being dependent
 of the explicit implementation of the cut-off)

If $D=4$ the situation is more complicated and interesting. The poles in 
(\ref{Gammagauge})
at $D=4$  have to  be interpreted as infrared logarithms containing in their argument
the box operator (as well as the mass for the matter contribution). They
  provide the well-known logarithmic running of the gauge couplings.
In this case the  bound on the gauge coupling is scale dependent and reads:
\bea
4 \pi \alpha_g(q) &\sim& [g_{0}^{-2}  + 
(\beta_0  + \beta_{1/2}) \log(\Lambda^2/(q^2 + m^2)) - 
\beta_1 \log(\Lambda^2/(q^2)]^{-1}   \nonumber \\
&\le& 
\left[(\beta_0  + \beta_{1/2}) \log(\Lambda^2/(q^2 + m^2)) - 
\beta_1 \log(\Lambda^2/q^2)\right]^{-1} ~~, ~~ D = 4 \;,
\label{alphab4}
\eea
 and, again generically, the limit is reached at infinite bare coupling. 
The physical meaning
of (\ref{alphab4}) is quite clear. Above the matter scale $m$ 
the gauge theory is weakly coupled at large $N$. However, below that scale, the
matter fields do no longer contribute to the running and, for the non-abelian
case, the theory evolves
towards strong coupling in the IR. It is quite easy to estimate the confinement scale of the
theory at infinite bare coupling,
\beq
{\Lambda_{conf} \over \Lambda} =  
\left( {m \over \Lambda} \right)^{1 + \beta_{tot}/\beta_1} ~ ,
\label{confscale} 
\eeq
where $\beta_{tot} = (\beta_0  + \beta_{1/2}) - \beta_1$ is the total
one-loop $\beta$-function coefficient
 (which is positive and large in the limit we consider).
Note that eq. (\ref{confscale}) can lead easily to exponentially
 large scale hierarchies. It would be interesting to try and verify (\ref{confscale})
by numerical simulations on scalar QCD with a small gauge group (say $SU(2)$) and
a large number of flavours, and work along these lines is in progress.
Note also that the logarithmic terms appearing in (\ref{alphab4}) 
should be actually accompanied by ''threshold" corrections, i.e. by a
scale-independent renormalization that would also affect the value of
the renormalized gauge coupling at the cut-off. Like the terms appearing in
(\ref{alphabound}),
 these finite corrections depend on
  the actual implementation of the cut-off, but, quite generically, we  expect
their contribution to $ \alpha_g^{-1}(\Lambda)$
to be $O(\beta_0  \pm \beta_{1/2} \pm 
\beta_1)$.

We now turn to the renormalization of the Newton constant. Here the situation
is the same at all $D \ge 4$. The low-energy action is local and, since 
 gravity couples equally to all fields, the constants $c_0, c_{1/2}$ are just
some calculable $N$-independent  numbers. Since the quantity
$(c_0 N_0 + c_{1/2} N_{1/2})$ is assumed to be parametrically large, graviton-loop
contributions are subleading at large $N$. The gauge field contribution, given the
fact that it is dominated by large virtual momenta, can be estimated accurately
in the one-loop approximation so that, the final result is:
\beq
\Gamma_{eff}^{gravity} = - {1 \over 2} \int \sqrt{-g} 
\left[  \kappa_{0}^{-2} + (c_0 N_0 + c_{1/2} N_{1/2} + c_1 N_1) \Lambda^{D-2}
  \right] R \;,
\label{Gammagrav}
\eeq
where $N_1$ is the total number of gauge bosons (the dimensionality of the
adjoint representation) and $c_1$ is also a calculable number of $O(1)$.

We thus obtain, at all scales, the bound on the effective Newton constant
\footnote {Similar statements on the large-$N$ limit of gravity
were made long ago by Tomboulis \cite{TO} and, more recently,
 were used in the context of black-hole
physics by several authors \cite{Lenny}.}
\beq
8 \pi \alpha_G \equiv \kappa^{2} \Lambda^{D-2} \sim
 [\kappa_{0}^{-2} \Lambda^{2-D} + 
(c_0 N_0 + c_{1/2} N_{1/2} + c_1 N_1)]^{-1} \le
 (c_0 N_0 + c_{1/2} N_{1/2} + c_1 N_1)^{-1} ,
\label{GNbound}
\eeq
where, once more, the bound is saturated in the compositeness limit, $\kappa_{0}
 \rightarrow \infty$.

We can finally rewrite both bounds (\ref{alphabound}) and (\ref{GNbound}) in 
the following suggestive way:
\beq
\alpha_g \le ~~ <N_f \, C_M ~ \frac{d_M}{d_A} + C_A>^{-1} ~~,
 \alpha_G  \le ~~ <N_f ~ d_M + d_A>^{-1}
\label{suggestive}
\eeq
where $<\dots>$ denotes an average over all matter fields.

Before turning our attention to physical applications of our claims we have 
to add a word of caution about the dependence of the bounds from the details
of the UV cutoff. Since the effects that we are interested in come from large virtual
momenta in the loops, it is clear that the UV completion of the model can influence
the final result. Work is in progress \cite{KV} to verify under which conditions
the claims are supported by explicit superstring theory calculations.

\section{Possible physical applications}

Let us assume that the results obtained
in the two preceeding Sections extend in the presence
 of matter self-interactions, leaving a detailed analysis of this
assumption  to future work, and
discuss  some possible physical applications.

The first  is  related to recent discussions
of entropy bounds. It is well known (see e.g. \cite{Wald})
that holographic bounds on entropy, as well as any other bounds involving
the Planck length are easily threatened in the presence of a large
number $N$ of species. This is because
 those bounds, being themselves geometrical, do not depend {\it explicitly} on
$N$, while the actual value of the entropy does increase with $N$.
However, since the bounds scale like an inverse power of
 the Planck length,  they can be saved if the latter decreases sufficiently fast
with $N$.
 Let us consider, in particular, the example of the 
Causal Entropy Bound (CEB) \cite{CEB}, recently discussed \cite{BFV}
in connection with explicit, small-coupling, large-temperature calculations
in CFTs \cite{Kutasov}. It was found \cite{BFV}
 that CFT entropy both at small  and, for CFTs with AdS duals,
at strong coupling,  fulfills CEB provided:

\begin{eqnarray}
\label{CEBcond}
\left(\frac{T}{M_P}\right)^{(D-2)}<  \frac{c}{N}
\end{eqnarray}

Clearly, if the maximal temperature is identified with $M_P$,
this inequality is in trouble at large $N$. However, in a cut-off theory of
gravity, the maximal temperature should rather be identified
with the UV cutoff $\Lambda$ (Cf. Hagedorn's temperature in string theory)
and, in that case, the inequality (\ref{CEBcond})
just becomes, at any $D$, our bound (\ref{bound}).
The bound (\ref{bound}) also avoids the potential
 problems, pointed out in ref. \cite{Ramy},
 with gravitational
 instability against formation of black holes from quantum fluctuations
in finite-size regions. 

The second possible use of the bounds (\ref{bound})  concerns
the dilaton stabilization and GUT-scale problems in string theory.
At tree-level, the gauge coupling and the ratio of the string (cutoff)
scale to the Planck mass are given by the expectation value of a scalar field, 
the dilaton. Perturbatively, the dilaton's VEV is
undetermined, since the dilaton acquires no potential/mass. This leads both to
possible large violations of the EP \cite{EP}, and to a possibly large space-time
dependence of fundamental constants, such as $\alpha$. Furthermore, the ratio
of the GUT scale to the Planck mass tends to come out too large \cite{SU}, 
in perturbative heterotic theory.
The first two problems can be solved if  non-perturbative effects induce a
  dilaton potential 
that provides the dilaton with a mass 
and freezes it together with the gauge and gravitational coupling. The problem
with GUT and Planck scales
 can be probably solved \cite{Witten} by going to strong string coupling 
(i.e. to 11-dimensional
supergravity) while keeping the $4D$ couplings small.

It looks highly improbable, however, that dilaton stabilization
can be achieved at weak $4D$  coupling, since any non-perturbative potential
falls to zero asymptotically (at zero coupling). Strong-weak $S-$duality 
makes stabilization very unlikely also 
in cases where a strong $4D$  coupling regime can be
replaced by a weak coupling one in a dual description. 
It looks therefore
that  dilaton stabilization may occur either  near the self-dual value,
 ($S\sim 1$), where S-duality is of no help \cite{BdA}, or at strong 
bare $4D$  coupling, if this case cannot be mapped into a 
weak-coupling situation. 
 At first sight, however, either solution
of the dilaton stabilization problem would lead to unphysical values 
for the unified gauge coupling and for the ratio $M_P/M_{GUT}$.

Our point  is that these  problem 
 can be avoided if the tree-level value of $M_P/M_s $ is renormalized
down to a value $O(1/N)$ by loop effects. Similarly, 
 $\alpha_{GUT}$ gets renormalized
downward, from  a large tree-level value to $\alpha_{GUT} \sim
 <N_f \, C_M ~ \frac{d_M}{d_A} + C_A> ^{-1}$.
For large-rank gauge groups (like $E_6$) and/or large matter
representations this could easily give an acceptable
value for the renormalized gauge coupling. Furthermore, given the fact that,
typically, $<N_f \, C_M ~ \frac{d_M}{d_A} + C_A> / 
<N_f ~ d_M + d_A> \sim (C_A, C_M) /  d_A $
one would find, as an order-of-magnitude relation,
\begin{eqnarray}
\label{Scales}
M_P^2/M_{GUT}^2 \sim N \sim \alpha_{GUT}^{-2} \; ,
\end{eqnarray}
in close similarity with the open-string relation that is known to
provide a much better agreement between $M_{GUT}$ and $M_s$.
Finally, a decoupling mechanism similar to the one proposed
in \cite{DP} would be operational near these non-trivial minima of the dilaton
potential. Notice that, form this point of view, a self-dual value of $S$ or an
infinite bare coupling ($S \rightarrow 0$) give essentially the same physics.

As a final application of this picture, I will mention the recently investigated 
possibility that a runaway dilaton may play the role of
quintessence \cite{GPV}. For this to work we have to assume that,
besides automatically decoupling from baryonic matter in the strong-coupling limit, 
the dilaton retains, in the same limit, a non-trivial
coupling to dark matter. The model naturally leads \cite {DPV} 
to acceptable, though not necessarily unobservable, 
violations of the equivalence principle (in particular 
of universality of free fall) and to tiny (and probably unobservable) 
variations of the 
fundamental ``constants".

\section{Acknowledgements}
I am grateful to C. Bachas, R. Brustein, T. Damour, M. Gasperini, E. Kohlprath,
 C. Kounnas, G. Marchesini, K. Meissner, F. Piazza,
E. Rabinovici, G.C. Rossi, A. Schwimmer and E. T. Tomboulis
 for useful discussions and/or correspondence over several years.
I also   wish to
acknowledge the support of a ``Chaire Internationale Blaise
Pascal", administered by the ``Fondation de L'Ecole Normale
Sup\'erieure'', during most of this work.

\end{document}